\documentclass[10pt,letterpaper,english]{article}
\usepackage{mathptmx}
\usepackage[T1]{fontenc}
\usepackage[utf8]{inputenc}
\usepackage[final]{microtype}
\usepackage{amsmath,amssymb,mathrsfs}
\usepackage{graphicx}
\usepackage{afterpage}
\usepackage{amsthm}
\theoremstyle{definition}

\numberwithin{equation}{section}
\usepackage{setspace}
\usepackage{needspace}
\usepackage[provide=*,english]{babel}
\usepackage{xcolor}
\definecolor{ReferenceGreen}{RGB}{0,128,0}
\definecolor{ReferenceRed}{RGB}{204,0,0}
\usepackage{hyperref}
\hypersetup{
  colorlinks=true,
  linkcolor=ReferenceRed,
  citecolor=ReferenceGreen,
  urlcolor=ReferenceGreen,
  filecolor=ReferenceGreen,
  pdfborder={0 0 0},
  bookmarksnumbered=true,
  bookmarksopen=false,
  linktoc=all,
  pdftitle={Blanket theory for dense networks},
  pdfauthor={Simone Franchini},
  pdfsubject={Constructive cavity method for dense networks with non-Gaussian noise},
  pdfkeywords={Blanket theory, dense networks, magnetization, cavity method, free energy}
}

\AtBeginDocument{
  \frenchspacing
  \setlength{\abovedisplayskip}{12pt}
  \setlength{\belowdisplayskip}{12pt}
  \setlength{\abovedisplayshortskip}{12pt}
  \setlength{\belowdisplayshortskip}{12pt}
}

\title{Blanket theory for dense networks}
\author{Simone Franchini}
\date{}
\begin{document}
\maketitle
\begin{center}
\textit{\small CNR-ISTC, Via Gian Domenico Romagnosi 18, 00196 Rome, Italy}
\end{center}
\vspace{14pt}

\begin{abstract}
\setstretch{1.1}
We show how Blanket theory provides a constructive method for calculating cavity functionals in dense spin networks. Building on the description through interfaces and magnetization shells, we develop the calculation from the low-temperature fluctuations around the ground state. In this regime, dilute defects reveal the mechanism of random energy model (REM) universality for linear energies, while their sharing and renewal across layers organize the energies and Gibbs weights into a hierarchy. Because the same configurations determine the covariances of the fresh cavity fields, this construction identifies the fields and weights with an Aizenman--Sims--Starr random overlap structure. The cavity contributions can then be evaluated through recursive fractional means, yielding an explicit free-energy functional parametrized by layer magnetizations and hierarchical exponents, with coefficients determined by the interaction variances and the network normalization. Under the stated moment and normalization conditions, comparison of pressures transfers the limiting mean pressure from the Gaussian representative to networks with independent centered non-Gaussian couplings having the same variances. We illustrate the method with the deterministic Curie--Weiss model, recover the Parisi functional for the Sherrington--Kirkpatrick model, and derive the functionals for restricted and deep Boltzmann machines, including the full replica symmetry breaking (fullRSB) limit.
\end{abstract}
\clearpage
\begingroup
\setstretch{1}
\fontsize{10}{12}\selectfont
\setcounter{tocdepth}{2}
\tableofcontents
\endgroup
\clearpage

\section{Introduction}
Blanket theory~\cite{SimplifiedParisi,RSBWithoutReplicas,REMUniversality,BlanketTheory,Multispin,SPA3} describes spin networks by organizing their interactions according to a partition of the sites. Each layer contributes a core of internal couplings and an interface connecting it to the preceding layers. In dense systems, refinement makes the internal contribution negligible under the appropriate scaling conditions, leading to a thermodynamic description through interfaces and states of fixed magnetization. The purpose of the present work is to show how this description can be used to construct and calculate cavity functionals. We thus continue the development of Ref.~\cite{BlanketTheory} by making explicit the connection between the decomposition of the interactions and the calculation of free energy.

The constructive cavity method (CCM) uses the layer decomposition to calculate the Parisi functional for the Sherrington--Kirkpatrick (SK) model~\cite{SimplifiedParisi}. The kernel representation makes the organization of the states and their weights explicit~\cite{RSBWithoutReplicas}, while the description of low-temperature excitations around the ground state provides the physical justification for the construction~\cite{SPA3}. We follow that mechanism here. Once the core contribution has been controlled, rectification on the ground state describes these excitations as a dilute gas of defects, whose number in each layer is fixed by its magnetization. Controlling the term bilinear in the defects reduces the fluctuations to linear energies and reveals the mechanism of random energy model (REM) universality~\cite{REMUniversality,REMLinear,REMLinearLetter}. Descendants inherit the defects and energy contributions of their ancestors, while new branches introduce new fluctuations. This organization provides the physical basis for the hierarchical description of the energies and Gibbs weights, developed first with finitely many defects and then extended to exponential populations of states.

The usefulness of this hierarchy for cavity calculations comes from its joint description of the weights and fields. Counting states with energies near the edge determines the exponents of the hierarchical masses; their Poisson representation~\cite{RuelleCascades} then makes averages over these masses calculable through fractional means. The same defect configurations determine the covariances of the fresh cavity fields, allowing the constructive variables and Gibbs weights to be identified with the random overlap structure (ROSt) of Aizenman--Sims--Starr~\cite{AizenmanSimsStarrOriginal,AizenmanSimsStarr}. The contribution of the added spin and the correction due to the change in reservoir normalization are therefore evaluated on the same reservoir by applying the fractional-mean identity recursively. This gives the functional~\eqref{eq:functional}, parametrized by layer magnetizations and hierarchical exponents. The common component of the linear field remains in the terminal function and is integrated after the hierarchical averages. Averaging the cavity contributions along a growing family of networks connects this calculation to the thermodynamic pressure.

To apply the construction to heterogeneous networks, the geometry of the states must retain the different strengths of the interactions. We therefore extend the population estimates to sums weighted by the coupling variances, so that the same configurations determine the covariances required by the cavity calculation. The model allows independent centered couplings with different laws and variances, including absent interactions, and keeps the normalization of Blanket theory explicit. Under the moment and scaling conditions stated in Subsection~\ref{subsec:model}, replacing the couplings successively by Gaussian variables with the same variances changes the mean pressure per spin by a vanishing amount~\cite{CarmonaHu,ChatterjeeLindeberg}. This comparison is performed at fixed temperature, before ground-state rectification, and transfers the thermodynamic value calculated with the Gaussian representative to the original network.

Sections~\ref{sec:blanket}--\ref{sec:cavity} develop this passage from the interaction decomposition to the explicit functional. The examples in Section~\ref{sec:examples} show how the resulting method adapts to different networks. Curie--Weiss provides the elementary deterministic case, in which the energy is constant on each magnetization shell and the cavity contribution can be calculated directly. For SK, the same construction recovers the canonical Parisi functional. Restricted and deep Boltzmann machines~\cite{CCMBoltzmannMachines,ChenIssaMourrat2026,BoursierMourrat2026} then show how the interaction architecture enters the calculation: intersecting the Blanket partition with the network layers determines the relevant magnetizations and fields, yielding the corresponding functionals and their full replica symmetry breaking (fullRSB) limit. Across these examples, the interactions determine the coefficients while the recursive averaging procedure supplies a common method of calculation.

\section{Hamiltonian, partition, and defects}
\label{sec:blanket}
We begin with the interaction decomposition of Blanket theory~\cite{SimplifiedParisi,RSBWithoutReplicas,REMUniversality,BlanketTheory,Multispin,SPA3}, which provides the starting point for the constructive calculation. After specifying the model, its normalization, and the pressure comparison for non-Gaussian couplings, we control the core and describe the remaining interface energy through magnetization shells and defects relative to the ground state~\cite{SPA3}. The bound on the bilinear defect term then isolates the linear fluctuations whose statistics are studied in Section~\ref{sec:hierarchy}.

\subsection{General model and normalization}
\label{subsec:model}
We consider a set of sites and the corresponding Ising configurations:
\begin{equation}
 V=\{i:i\leq N\},\qquad \sigma_V=\{\sigma_i:i\in V\}\in\Omega^V=\{-1,1\}^V.
 \label{eq:sites}
\end{equation}
The site indices are positive integers. Following the convention of Ref.~\cite[Eqs.~(16), (23)]{BlanketTheory}, the Hamiltonian is
\begin{equation}
 H(\sigma_V)=\sum_{i,j\in V}H_{ij}\sigma_i\sigma_j.
 \label{eq:model}
\end{equation}
The uniform normalization of the interaction matrix is
\begin{equation}
 H_{ij}=\frac{\beta J_{ij}}{g(V^2)}.
 \label{eq:normalization}
\end{equation}
Here the inverse temperature is already included in the coefficients. The positive normalization function depends on the volume of the network; its choice must make the free energy extensive. The couplings may have different laws and different variances, and may be zero. The variance of the normalized coefficient is
\begin{equation}
 \Gamma_{ij}:=\operatorname{Var}(H_{ij})=\frac{\beta^2}{g(V^2)^2}\operatorname{Var}(J_{ij}).
 \label{eq:variance}
\end{equation}
The two orientations of a pair are counted separately. For the derivation with noise, we consider independent centered unnormalized coefficients, with uniformly bounded variances and finite third moments. The variance matrix remains arbitrary within these conditions. No assumption of homogeneity among sites is required in the decomposition. In Subsection~\ref{subsec:CW} we also treat the deterministic case with constant interaction directly, retaining its mean value.

The partition function and the Gibbs measure are, respectively,
\begin{equation}
 Z=\sum_{\sigma_V\in\Omega^V}e^{-H(\sigma_V)},\qquad
 \mu(\sigma_V)=\frac{e^{-H(\sigma_V)}}{Z}.
 \label{eq:gibbs}
\end{equation}
We retain the convention of Ref.~\cite[Eq.~(22)]{BlanketTheory} for the dimensionless free energy:
\begin{equation}
 f=-\frac1N\log Z.
 \label{eq:free-energy}
\end{equation}
The pressure is its negative; the physical free energy is obtained by dividing the dimensionless free energy by the inverse temperature. The expectation denotes the average over the disorder. Spin averages are denoted by angle brackets, with the measure or the magnetization constraint specified in the subscript.

To treat non-Gaussian noise, we compare the pressure of the network with that of the model with Gaussian couplings having the same variances. The replacement of one coupling at a time~\cite{ChatterjeeLindeberg} applies to the full partition function before alignment with the ground state and the choice of fibers. With all other couplings fixed, regard the dimensionless free energy in Eq.~\eqref{eq:free-energy} as a function of the coefficient being replaced.
The derivatives of the logarithm are thermal cumulants of the product of the two spins. Since this product has absolute value one, the third cumulant has absolute value at most eight and
\begin{equation}
 \left|\frac{\mathrm d^3}{\mathrm dJ_{ij}^3}f(J_{ij})\right|
 \leq\frac{8\beta^3}{N\,g(V^2)^3}.
 \label{eq:replacement-third-derivative}
\end{equation}
We expand around zero to second order, with a third-order remainder. The constant term cancels; the linear and quadratic terms have the same expectation for the original coupling and for a centered Gaussian with the same variance. With $\tilde J_{ij}$ standard Gaussian and independent of the other couplings, it follows that
\begin{equation}
 \left|\mathbb E f(J_{ij})- \mathbb E f\!\left(\sqrt{\operatorname{Var}(J_{ij})}\,\tilde J_{ij}\right)\right| \leq\frac{C\beta^3}{N\,g(V^2)^3} \left[\mathbb E|J_{ij}|^3+ \bigl(\operatorname{Var}(J_{ij})\bigr)^{3/2}\right].
 \label{eq:replacement-one-bond-bound}
\end{equation}
The expectations in this step integrate over the replaced coefficient. The bound is uniform in the couplings held fixed.

By successively replacing all the couplings and summing the differences, we obtain the following bound on the difference between the mean pressures per spin:
\begin{equation}
 \frac{C\beta^3}{N g(V^2)^3}\sum_{i,j\in V}
 \left\{\mathbb E|J_{ij}|^3+[\operatorname{Var}(J_{ij})]^{3/2}\right\}\longrightarrow0.
 \label{eq:comparison}
\end{equation}
The constant is independent of the volume. We require the indicated remainder to tend to zero; for uniformly bounded third moments, it suffices that the volume divided by the cube of the normalization tend to zero. The comparison holds at fixed temperature and precedes any zero-temperature limit.

In what follows, we carry out the construction on the Gaussian representative, retaining the notation of~\eqref{eq:model}--\eqref{eq:variance}. Equation~\eqref{eq:comparison} transfers the limiting value of the pressure to the original network. For the family under study, we assume the existence of this limit and a normalization for which the subsequent energy bounds are negligible in the indicated limits.

\subsection{Blanket decomposition and interface dominance}
\label{subsec:partition}
Independently of the disorder, we choose an ordered partition into disjoint nonempty layers~\cite{RSBWithoutReplicas,BlanketTheory}:
\begin{equation}
 \mathcal V=\{V_\ell:\ell\in\Lambda\},\qquad V=\bigcup_{\ell\in\Lambda}V_\ell.
 \label{eq:partition}
\end{equation}
The layer indices in the following sums belong to the sets
\begin{equation}
 \Lambda=\{1,\ldots,L\},\qquad \Lambda_\ell=\{1,\ldots,\ell\},\qquad \Lambda_0=\varnothing.
 \label{eq:indices}
\end{equation}
The cardinality of a layer and the cumulative sets are defined by
\begin{equation}
 N_\ell=|V_\ell|,
 \label{eq:layersize}
\end{equation}
\begin{equation}
 Q_\ell=\bigcup_{t\in\Lambda_\ell}V_t,\qquad Q_0=\varnothing.
 \label{eq:cumulative}
\end{equation}
The corresponding volume fractions satisfy
\begin{equation}
 q_\ell=\frac{|Q_\ell|}{N},\qquad p_\ell=\frac{N_\ell}{N}=q_\ell-q_{\ell-1}.
 \label{eq:fractions}
\end{equation}
The blanket collects the pairs introduced when the layer is added:
\begin{equation}
 W_\ell=Q_\ell^2\setminus Q_{\ell-1}^2.
 \label{eq:volumes}
\end{equation}
Its internal part is the core; the two rectangular parts form the interface:
\begin{equation}
 \widehat W_\ell=V_\ell^2,\qquad
 \overline W_\ell=(V_\ell\times Q_{\ell-1})\cup(Q_{\ell-1}\times V_\ell).
 \label{eq:core-interface-sets}
\end{equation}
As in Ref.~\cite[Eq.~(64)]{BlanketTheory}, the normalization ratios are
\begin{equation}
 \widehat\gamma_\ell=\frac{g(V_\ell^2)}{g(V^2)},\qquad
 \overline\gamma_\ell=\frac{g(\overline W_\ell)}{g(V^2)}.
 \label{eq:normalization-ratios}
\end{equation}
The interface ratio is used for the layers after the first, where the interface is nonempty.
The core Hamiltonian is obtained by multiplying that of the layer, normalized on its own set, by the first ratio:
\begin{equation}
 \widehat H_\ell(\sigma_{V_\ell})=\widehat\gamma_\ell H(\sigma_{V_\ell}).
 \label{eq:core-local}
\end{equation}
For the layers after the first, the incoming field includes both orientations:
\begin{equation}
 h_i(\sigma_{Q_{\ell-1}})=\frac{\beta}{g(\overline W_\ell)}
 \sum_{j\in Q_{\ell-1}}(J_{ij}+J_{ji})\sigma_j,\qquad i\in V_\ell.
 \label{eq:interface-field}
\end{equation}
The interface energy is therefore
\begin{equation}
 \overline H_\ell(\sigma_{V_\ell}\mid\sigma_{Q_{\ell-1}})
 =\overline\gamma_\ell\sum_{i\in V_\ell}\sigma_i h_i(\sigma_{Q_{\ell-1}}).
 \label{eq:interface-local}
\end{equation}
The first layer has an empty interface and zero contribution. The sum of the interfaces defines the IO Hamiltonian:
\begin{equation}
 \bar H(\sigma_V)=\frac{\beta}{g(V^2)}\sum_{\ell\in\Lambda}\;\sum_{t<\ell}\;
 \sum_{i\in V_\ell,j\in V_t}(J_{ij}+J_{ji})\sigma_i\sigma_j.
 \label{eq:IO}
\end{equation}
Figure~\ref{fig:cores} shows the partition of the interactions.\par
\afterpage{\clearpage
\begin{figure}[p]
\centering
\includegraphics[width=\textwidth,height=.62\textheight,keepaspectratio]{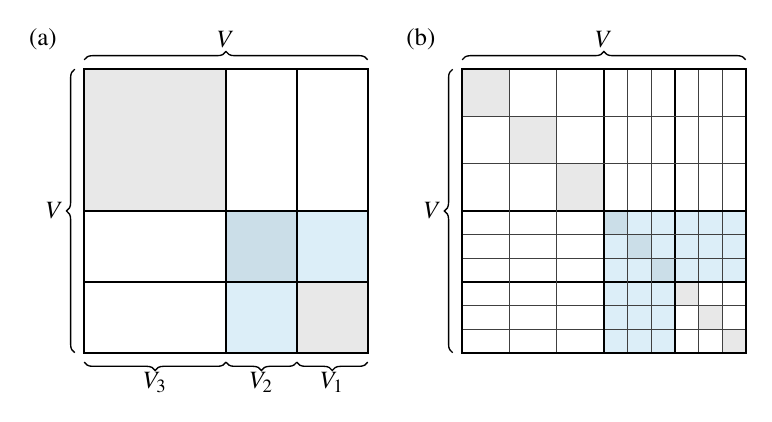}
\begingroup\small\setstretch{1}
\caption{Blanket partition of the interaction matrix~\cite{RSBWithoutReplicas,BlanketTheory}. The highlighted sector is a blanket from~\eqref{eq:volumes}: it comprises the square internal to the layer and the two orientations of the interface with the preceding layers, defined in~\eqref{eq:core-interface-sets}. The IO Hamiltonian retains the interfaces. Refinement reduces the number of pairs in the cores, and~\eqref{eq:core} bounds the corresponding error in the pressure per spin.}
\label{fig:cores}
\endgroup
\end{figure}
\clearpage}
 Counting the internal pairs and bounding the Gaussian maximum give
\begin{equation}
 \frac1N\mathbb E\sup_{\sigma_V}|H(\sigma_V)-\bar H(\sigma_V)|
 \leq C\frac{\beta\sqrt N}{g(V^2)}\sqrt{\sum_{\ell\in\Lambda}p_\ell^2}.
 \label{eq:core}
\end{equation}
The bound uses the law of the Gaussian representative before alignment, with independent centered couplings and uniformly bounded variances. The constants are independent of the volume and of the deterministic partition. For a fixed configuration, the core variance is at most
\begin{equation}
 \frac{C\beta^2}{g(V^2)^2}\sum_{\ell\in\Lambda}N_\ell^2.
 \label{eq:core-variance-bound}
\end{equation}
The maximum is taken over the Ising configurations: the union bound for the Gaussian tails and their integration give~\eqref{eq:core}.
Comparing the Boltzmann sums preserves the multiplicities and bounds the difference between the free energies per spin by the maximum energy per spin.

If the ratio of the square root of the volume to the normalization remains bounded, the core becomes negligible at fixed temperature as the maximum relative volume of the layers tends to zero.

\subsection{Ground-state rectification and linear reduction}
\label{subsec:defects}
We write the model directly in coordinates aligned with a ground state of the IO Hamiltonian, retaining the notation for spins, coefficients, the Hamiltonian, and the Gibbs measure~\cite{SPA3}. The reference is the configuration in which all spins are positive, denoted by $1_V$:
\begin{equation}
 \bar H(1_V)=\min_{\sigma_V\in\Omega^V}\bar H(\sigma_V).
 \label{eq:groundstate}
\end{equation}
Alignment preserves the energy and the partition function. The assumptions on the law of the couplings and the variances in Subsection~\ref{subsec:model} continue to refer to the coefficients before alignment; this convention also applies to the subsequent formulas. The aligned coefficients have the law induced by the choice of the ground state. Global inversion produces the equivalent sector, whose factor of two is negligible in the pressure per spin. The construction describes the low-temperature regime in which the gas of defects relative to the reference is dilute.

The defects in the layer are
\begin{equation}
 X(\sigma_{V_\ell})=\{i\in V_\ell:\sigma_i=-1\}.
 \label{eq:defectsupport}
\end{equation}
Their density determines the magnetization:
\begin{equation}
 \epsilon_\ell=\frac{|X(\sigma_{V_\ell})|}{N_\ell},\qquad m_\ell=1-2\epsilon_\ell.
 \label{eq:defects}
\end{equation}
For the full profile, we use the set notation
\begin{equation}
 m_\Lambda=\{m_\ell:\ell\in\Lambda\}.
 \label{eq:magnetization-profile}
\end{equation}
The shell of the layer and the fiber associated with the profile are
\begin{equation}
 \Omega^{V_\ell}(m_\ell)=\left\{\sigma_{V_\ell}:N_\ell^{-1}\sum_{i\in V_\ell}\sigma_i=m_\ell\right\},
 \label{eq:shell}
\end{equation}
\begin{equation}
 \Omega^V(m_\Lambda)=\prod_{\ell\in\Lambda}\Omega^{V_\ell}(m_\ell).
 \label{eq:fibre}
\end{equation}
The uniform average over this fiber is denoted by angle brackets, with the profile in the subscript. The number of configurations is
\begin{equation}
 |\Omega^V(m_\Lambda)|=\prod_{\ell\in\Lambda}\binom{N_\ell}{\epsilon_\ell N_\ell}.
 \label{eq:shellcount}
\end{equation}
The IO partition function has the exact disintegration
\begin{equation}
 \bar Z=\sum_{m_\Lambda}|\Omega^V(m_\Lambda)|
 \left\langle e^{-\bar H(\sigma_V)}\right\rangle_{m_\Lambda}.
 \label{eq:disintegration}
\end{equation}
The sum runs over the admissible profiles. Applied also to the numerator of the average of an observable, the same identity preserves the Gibbs weights. For a fixed number of layers, the number of profiles is polynomial in the volume: the comparison between fibers includes their entropy and the full energy.

The center of the fiber is
\begin{equation}
 \left\langle\bar H(\sigma_V)\right\rangle_{m_\Lambda}
 =\frac{\beta}{g(V^2)}\sum_{\ell\in\Lambda}\;\sum_{t<\ell}m_\ell m_t
 \sum_{i\in V_\ell,j\in V_t}(J_{ij}+J_{ji}).
 \label{eq:center}
\end{equation}
Expanding the spins around the magnetizations gives
\begin{equation}
\begin{aligned}
 \bar H(\sigma_V)-\left\langle\bar H(\sigma_V)\right\rangle_{m_\Lambda}
 ={}&\frac{\beta}{g(V^2)}\sum_{\ell\in\Lambda}\,\sum_{i\in V_\ell}(\sigma_i-m_\ell)
 \left[\sum_{t\ne\ell}m_t\sum_{j\in V_t}(J_{ij}+J_{ji})\right]\\
 &+\frac{\beta}{g(V^2)}\sum_{\ell\in\Lambda}\,\sum_{t<\ell}\,
 \sum_{i\in V_\ell,j\in V_t}(J_{ij}+J_{ji})(\sigma_i-m_\ell)(\sigma_j-m_t).
\end{aligned}
\label{eq:linear-bilinear}
\end{equation}
For fixed disorder and profile, the linear coefficients are common to all configurations of the layer and include the mean contributions of all the other layers. The bilinear term satisfies the bound of Ref.~\cite{SPA3} with the normalization~\eqref{eq:normalization}:
\begin{equation}
\begin{aligned}
 &\frac1N\mathbb E\max_{\sigma_V\in\Omega^V(m_\Lambda)}
 \left|\frac{\beta}{g(V^2)}\sum_{\ell\in\Lambda}\;\sum_{t<\ell}\;
 \sum_{i\in V_\ell,j\in V_t}(J_{ij}+J_{ji})(\sigma_i-m_\ell)(\sigma_j-m_t)\right|\\
 &\qquad\leq C\frac{\beta\sqrt N}{g(V^2)}
 \bigl(\max_\ell\epsilon_\ell\bigr)^{3/2}\sqrt{\log\frac e{\max_\ell\epsilon_\ell}}.
\end{aligned}
\label{eq:bilinear}
\end{equation}
For the bilinear term, we use the identity
\begin{equation}
 \sigma_i-m_\ell=-2\bigl[\mathbb I_{\{i\in X(\sigma_{V_\ell})\}}-\epsilon_\ell\bigr],\qquad i\in V_\ell.
 \label{eq:appendix-indicators}
\end{equation}
The indicator function equals one on the specified support and zero elsewhere. For supports and signs fixed before the disorder, the term with two indicators has variance at most
\begin{equation}
 \frac{C\beta^2N^2}{g(V^2)^2}\bigl(\max_\ell\epsilon_\ell\bigr)^2.
 \label{eq:bilinear-indicator-variance}
\end{equation}
The logarithm of the number of possible choices of the supports and of the signs on their sites is at most
\begin{equation}
 CN\max_\ell\epsilon_\ell\log\frac e{\max_\ell\epsilon_\ell}.
 \label{eq:bilinear-support-entropy}
\end{equation} The Gaussian maximum, divided by the volume, therefore gives~\eqref{eq:bilinear}.
Here the maximum density is positive and does not exceed one half; if it is zero, the term vanishes. To complete the calculation, the norm of the symmetric matrix of couplings between layers is unchanged by rectification and has expectation at most $C\sqrt N$. In the mixed terms, the norms of the indicator and density vectors are respectively at most
\begin{equation}
 \sqrt{N\max_\ell\epsilon_\ell},\qquad \sqrt N\max_\ell\epsilon_\ell.
 \label{eq:mixed-vector-norms}
\end{equation} This yields the same prefactor as in~\eqref{eq:bilinear}, without the logarithm; the term with two densities is of quadratic order. The maximum over the signs includes those chosen by the ground state.

Comparing the Boltzmann sums over the same configurations gives the same bound on the difference between the pressures. The center~\eqref{eq:center} remains in the weight of the fiber.

We first take the thermodynamic limit at fixed partition and fixed positive densities, with macroscopic layers; we then pass to the dilute regime and refinement, keeping the errors~\eqref{eq:core} and~\eqref{eq:bilinear} negligible. Magnetizations and populations specify the hierarchical representative; they are selected after the functional has been calculated.

\section{State hierarchy and averages over the weights}
\label{sec:hierarchy}
The dilution of defects reveals the REM mechanism of the linear sums; sharing ancestors organizes these contributions into a GREM~\cite{REMUniversality,SPA3}. We extend the construction to exponential populations and retain the weighted covariances needed for the general network. Counting the energies then determines the weights and the averaging identity used in the cavity.

\subsection{REM and GREM of the dilute gas}
\label{subsec:hierarchy}
We also use set notation for the state indices. A complete path is denoted by $\alpha_\Lambda$ and its prefix by
\begin{equation}
 \alpha_{\Lambda_\ell}=\{\alpha_t:t\in\Lambda_\ell\}.
 \label{eq:state-indices}
\end{equation}
The same convention applies to the other state indices. We denote the layer configuration and the complete configuration, respectively, by
\begin{equation}
 \sigma_{V_\ell}^{\alpha_{\smash[b]{\Lambda_\ell}}},\qquad \sigma_V^{\alpha_\Lambda}.
 \label{eq:prefix}
\end{equation}
We assign a sample from the layer's shell to each node; all descendants retain this configuration. At fixed disorder and magnetizations, the linear contribution in~\eqref{eq:linear-bilinear} is a sum over a fixed population of coefficients. Independent samples of supports therefore produce independent energies with the same law at the level. Descendants reuse the energy of their ancestors.

With a finite number of defects and nodes, the probability of accidental intersections between the new supports tends to zero as the layer grows. The sum over the defect costs thus realizes the REM mechanism~\cite{DerridaREM}, while the reuse of contributions along the prefix produces the GREM composition~\cite{DerridaGREM}. The argument in Ref.~\cite{SPA3} applies this construction to the entire strip of the kernel, including nodes with different parents. Figure~\ref{fig:kernel} shows the geometry of the states and their weights.\par
\afterpage{\clearpage
\begin{figure}[p]
\centering
\includegraphics[width=\textwidth,height=.62\textheight,keepaspectratio]{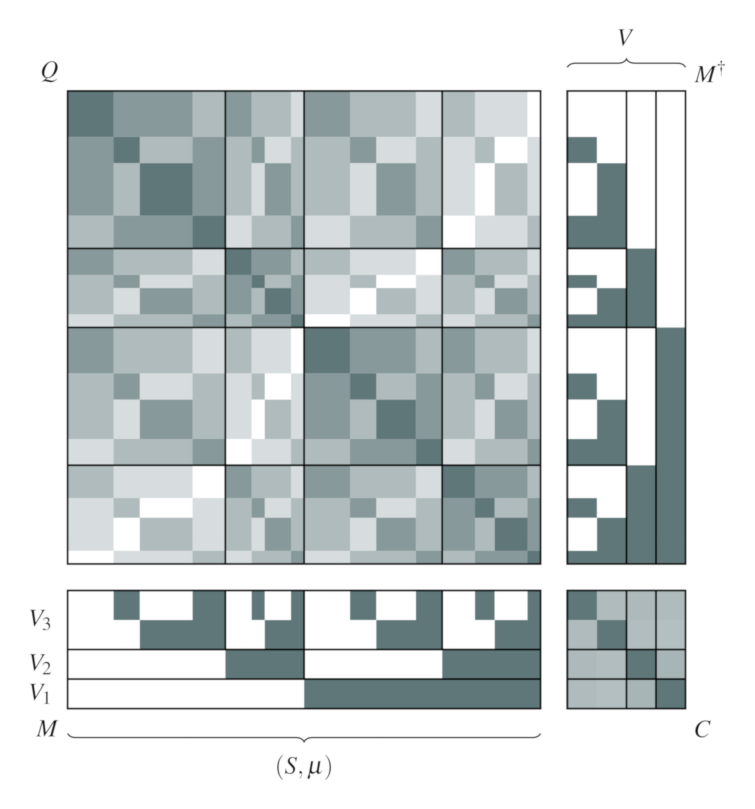}
\begingroup\small\setstretch{1}
\caption{Spin kernel, transpose, and associated matrices~\cite{RSBWithoutReplicas,SPA3}. The letters $M$ and $M^\dagger$ denote the kernel and its transpose, $Q$ the overlap matrix, $C$ the matrix of correlations between sites, and $S$ the set of states. The sites are ordered by layers; the widths of the states represent the Gibbs weights. The correlations between sites and the overlaps between states are obtained by integrating the product of two sections of the kernel over the other coordinate. The states in~\eqref{eq:prefix} share their ancestors' configurations as described in Subsection~\ref{subsec:hierarchy}. In the dilute gas, this sharing determines the hierarchical overlap values in~\eqref{eq:rho}.}
\label{fig:kernel}
\endgroup
\end{figure}
\clearpage}

At fixed positive densities, the shells have exponential capacity. If each node at the preceding level has $n_\ell$ children, the capacity imposes
\begin{equation}
 \prod_{t\in\Lambda_\ell}n_t\leq\binom{N_\ell}{\epsilon_\ell N_\ell}.
 \label{eq:capacity}
\end{equation}
The product counts all nodes in the strip. We denote by $\mathbb P$ the probability with respect to the independent and uniform sampling of supports on the shells. Two independent supports in the same layer have a hypergeometric intersection. For sufficiently small densities,
\begin{equation}
 \mathbb P\!\left\{|X(\sigma_{V_\ell}^{\alpha_{\smash[b]{\Lambda_\ell}}})\cap X(\sigma_{V_\ell}^{\beta_{\smash[b]{\Lambda_\ell}}})|
 \geq N_\ell\epsilon_\ell^{3/2}\right\}
 \leq(e\sqrt{\epsilon_\ell})^{N_\ell\epsilon_\ell^{3/2}}.
 \label{eq:appendix-intersection}
\end{equation}
The probability of the union of the events over all pairs of nodes in the strip tends to zero when
\begin{equation}
 \limsup_{N\to\infty}\frac2{N_\ell}\log\!\left(\prod_{t\in\Lambda_\ell}n_t\right)
 <\epsilon_\ell^{3/2}\log\frac1{e\sqrt{\epsilon_\ell}}.
 \label{eq:appendix-population}
\end{equation}
This allows exponential populations in all strips, with the number of layers and their positive fractions fixed. The bound covers nodes with different prefixes. The ratio of the intersection to the number of defects then tends to zero in the dilute regime.

The law of the energies is that of the complete sums in~\eqref{eq:linear-bilinear}. Rectification selects the coefficients through the ground state; the count of rare energies therefore uses this effective law. In layers with a nondegenerate field, the dispersion of the costs allows the energy to vary while the number of defects remains fixed. Together with the margin between the sampled population and the shell capacity, this motivates the regularity of the energy edge discussed in Ref.~\cite[Secs.~3.3--3.4]{SPA3}. Levels with zero energy increment are omitted from the energy count. The mechanism of REM universality for linear energies is developed in Refs.~\cite{REMUniversality,REMLinear,REMLinearLetter}. Earlier work includes the energy statistics of the number partitioning problem~\cite{MertensRandomCosts,BaukeFranzMertens}, the local results of Ref.~\cite{BovierKurkovaLocalREM}, and those for diluted populations in Ref.~\cite{BenArousGayrardKuptsov}.

This construction preserves the sharing and renewal of defects when passing to extensive populations. Equation~\eqref{eq:bilinear} controls the error of the linear reduction on the configurations of the fiber; the dilute gas provides the physical justification for the hierarchical representation, as in Ref.~\cite{SPA3}. Entropies, magnetizations, and level populations remain parameters of the description.

The physical overlap is
\begin{equation}
 q(\sigma_V,\tau_V)=\frac1N\sum_{i\in V}\sigma_i\tau_i.
 \label{eq:overlap}
\end{equation}
For two branches that separate after the indicated level, the configurations inherited from the ancestors coincide, whereas on subsequent layers the mean product of the two samples equals the square of the magnetization. The hierarchical values of Ref.~\cite{SPA3} are therefore
\begin{equation}
 \rho_\ell=q_\ell+\sum_{t>\ell}p_t m_t^2,\qquad \ell=0,\ldots,L.
 \label{eq:rho}
\end{equation}
The increments are
\begin{equation}
 \rho_\ell-\rho_{\ell-1}=p_\ell(1-m_\ell^2).
 \label{eq:rho-increments}
\end{equation}
The control of intersections bounds the deviation from the hierarchical overlap. The dependence on the last common ancestor expresses ultrametricity; in the dilute regime, the overlaps tend to one in the rectified sector.

For heterogeneous covariances, the weights must be retained. Concentration for sums sampled without replacement~\cite[Sec.~6]{Hoeffding}, applied to the bounded variances, gives, for every receiving site and positive threshold,
\begin{equation}
 \mathbb P\!\left\{\left|\frac1{N_\ell}\sum_{j\in V_\ell}\operatorname{Var}(J_{ij})
 \bigl[\mathbb I_{\{j\in X(\sigma_{V_\ell}^{\alpha_{\smash[b]{\Lambda_\ell}}})\}}-\epsilon_\ell\bigr]\right|>s\right\}\leq2\exp\!\left[-\frac{N_\ell s^2}{C\epsilon_\ell}\right].
\label{eq:appendix-weighted-linear}
\end{equation}
The same holds for the other orientation. We choose the tolerance
\begin{equation}
 s=C\epsilon_\ell^{5/4}\sqrt{\log\frac e{\epsilon_\ell}}=o(\epsilon_\ell).
 \label{eq:weighted-linear-tolerance}
\end{equation}
With a sufficiently large constant, the union of the tail events simultaneously controls all nodes and all receiving sites in the network. The exponent has the order of that in~\eqref{eq:appendix-population}; the number of sites adds only a polynomial factor. When the product of two spins is expanded in terms of their indicators,~\eqref{eq:appendix-intersection} also controls the weighted intersection, since the variances are bounded.

For the quadratic term, we fix two paths and their common prefix. In the following formula, the angle brackets denote the average over the supports of the two paths, sampled uniformly while preserving the common prefix. A swap in a support changes the normalized sum by at most $C/N$. The number of sampled sites to be revealed is at most
\begin{equation}
 2N\max_\ell\epsilon_\ell.
 \label{eq:weighted-quadratic-samples}
\end{equation} The bounded-differences inequality~\cite{Azuma} gives
\begin{equation}
\begin{aligned}
 &\mathbb P\!\left\{\left|\frac1{N^2}\sum_{i,j\in V}\operatorname{Var}(J_{ij})
 \left[\sigma_i^{\alpha_\Lambda}\sigma_i^{\beta_\Lambda}\sigma_j^{\alpha_\Lambda}\sigma_j^{\beta_\Lambda}
 -\left\langle\sigma_i^{\alpha_\Lambda}\sigma_i^{\beta_\Lambda}\sigma_j^{\alpha_\Lambda}\sigma_j^{\beta_\Lambda}\right\rangle_{m_\Lambda}\right]\right|>s\right\}\\
 &\qquad\leq2\exp\!\left[-\frac{Ns^2}{C\max_\ell\epsilon_\ell}\right].
\end{aligned}
\label{eq:appendix-weighted-quadratic}
\end{equation}
With the same choice of populations, all pairs are controlled by the tolerance
\begin{equation}
 s=C\bigl(\max_\ell\epsilon_\ell\bigr)^{5/4}\sqrt{\log\frac e{\max_\ell\epsilon_\ell}}.
 \label{eq:weighted-quadratic-tolerance}
\end{equation} The averages over the shells retain the diagonal and the finite corrections due to the fixed cardinality. Finally, the covariance prefactors are restored using~\eqref{eq:variance}: negligibility refers to the errors rescaled in this way. On the same event, any selection or reweighting of the sampled configurations, including Gibbs reweighting, preserves the uniform bounds. The effective covariances of the network thus remain available and will be calculated in Section~\ref{sec:cavity}.

\subsection{Hierarchical weights and fractional mean}
\label{subsec:means}
The cost of a given level energy is calculated from the full law of the linear contribution in~\eqref{eq:linear-bilinear}, at fixed disorder and profile. With the notation of Ref.~\cite[Sec.~3.4]{SPA3}, the transform is
\begin{equation}
\begin{aligned}
 I_\ell(e_\ell)=\sup_{s\in\mathbb R}\Bigg\{-s e_\ell
 -\lim_{N\to\infty}\frac1N\log\Bigg\langle\exp\Bigg[
 -\frac{s\beta}{g(V^2)}\sum_{i\in V_\ell}(\sigma_i-m_\ell)\\
 {}
 \times\sum_{t\ne\ell}m_t\sum_{j\in V_t}(J_{ij}+J_{ji})
 \Bigg]\Bigg\rangle_{m_\ell}\Bigg\}.
\end{aligned}
\label{eq:rate}
\end{equation}
The argument is the level energy per spin of the entire system; the average is calculated over the level's shell. We consider the regular lower branch of the cost when the indicated limit exists. The mean count of paths with assigned energies combines the population and the cost of the respective windows~\cite[Theorem~3.1]{JanaRaoLDP}. For each prefix, it requires
\begin{equation}
 \sum_{t\in\Lambda_\ell}I_t(e_t)
 \leq\lim_{N\to\infty}\frac1N\log\!\left(\prod_{t\in\Lambda_\ell}n_t\right).
 \label{eq:balance}
\end{equation}
The edge saturates the constraint. Consecutive levels that reach the edge together are grouped into an effective level: at the energy minimum for a fixed total cost, the slopes of their costs coincide. Groups with the smallest threshold are chosen successively, merging those with the same threshold, as in Ref.~\cite{SPA3}, which extends the grouping of GREM levels~\cite{DerridaGardnerGREM,BovierKurkova} to the law of the rectified costs.

In the condensed regime, we use the exponent of the effective level to denote the positive slope at the edge:
\begin{equation}
 -\partial_{e_\ell}I_\ell(e_\ell)=\lambda_\ell,\qquad 0<\lambda_\ell<1.
 \label{eq:slope}
\end{equation}
The formula concerns an effective level; before grouping, the same slope belongs to all grouped layers. The energy in Eq.~\eqref{eq:rate} already includes the temperature. Following the shell calculation of Ref.~\cite[Sec.~3.4]{SPA3}, exponential weighting fixes the mean energy and defect number together. In the regular regime considered there, dispersion of the rectified costs leaves energy fluctuations after the cardinality is fixed. For a single list, the joint fluctuation prefactor is of inverse-volume order; conditioning on the defect number leaves an inverse-square-root prefactor. Its relative variation under finite energy shifts vanishes, and its contribution is absorbed into a logarithmic correction to the centering, evaluated on the actual rectified coefficients. Grouped levels retain the internal prefix constraints. We denote by $E$ the order-one displacement from this centered edge. The finite-shift count ratio is exponential with the slope in Eq.~\eqref{eq:slope}; independent renewal of the rare samples gives the Poisson intensity
\begin{equation}
 \lambda_\ell e^{\lambda_\ell E}\,\mathrm dE,\qquad E\in\mathbb R.
 \label{eq:energyintensity}
\end{equation}
The Boltzmann weight is
\begin{equation}
 w=e^{-E}.
 \label{eq:energy-weight}
\end{equation}
The change of variable gives the intensity of the weights:
\begin{equation}
 \lambda_\ell w^{-1-\lambda_\ell}\,\mathrm dw,\qquad w>0.
 \label{eq:weight-intensity}
\end{equation}
In the condensed regime, the same shell calculation bounds the expected count in finite energy intervals uniformly in the volume and interval location. Above an expanding window, the count grows with an exponential slope below one, so multiplication by the Boltzmann weight makes the remaining total vanish. The probability of a state below the lower edge also vanishes. These controls extend the description of the extremes to the total and normalized weights~\cite[Sec.~3.4]{SPA3}, retaining all descendant totals in each complete prefix mass.

The condensed effective levels have ordered exponents; the final thermal levels enter with exponent one. The complete profile is
\begin{equation}
 \lambda_\Lambda=\{\lambda_\ell:\ell\in\Lambda\},\qquad
 0<\lambda_1\leq\cdots\leq\lambda_L\leq1.
 \label{eq:exponents}
\end{equation}
The ordering refers to the levels after grouping; the grouped layers can be retained by assigning them the same exponent.

For strictly increasing exponents between zero and one, each prefix carries an independent list with intensity~\eqref{eq:weight-intensity}. We use the weight notation of Ref.~\cite{CCMBoltzmannMachines}:
\begin{equation}
 \{w_\ell(\alpha_{\Lambda_\ell}):\alpha_\ell\geq1\}.
 \label{eq:PPP}
\end{equation}
The weights of complete paths are normalized globally:
\begin{equation}
 \bar\mu_{\mathrm{GREM}}(\sigma_V^{\alpha_\Lambda})=
 \frac{\prod_{\ell\in\Lambda}w_\ell(\alpha_{\Lambda_\ell})}
 {\sum_{\beta_\Lambda}\prod_{\ell\in\Lambda}w_\ell(\beta_{\Lambda_\ell})}.
 \label{eq:cascade}
\end{equation}
This is the Ruelle cascade~\cite{RuelleCascades}. The center~\eqref{eq:center} remains in the comparison between fibers in~\eqref{eq:disintegration}. Equal values of the exponents are treated by grouping the increments; the endpoints zero and one are obtained by continuity.

The fresh marks are independent and identically distributed across nodes, and independent of the list. For a positive function of these marks, with the necessary moments, rescaling the marked points gives
\begin{equation}
 \mathbb E\log\frac{\sum_{\alpha_\ell}w_\ell(\alpha_{\Lambda_\ell})
 \mathcal O(\tilde z_\ell^{\alpha_{\smash[b]{\Lambda_\ell}}})}
 {\sum_{\alpha_\ell}w_\ell(\alpha_{\Lambda_\ell})}
 =\frac1{\lambda_\ell}\log\mathbb E \mathcal O(\tilde z_\ell)^{\lambda_\ell}.
 \label{eq:meanidentity}
\end{equation}
The average includes the list and the marks, at fixed prefix. The marked total has the law of the original total multiplied by the fractional mean; subtracting the mean logarithms gives~\eqref{eq:meanidentity}. In the cascade, we proceed from the last level to the first, including the totals of the descendants. The stability of the lists under these factors yields the same identity with the exponent of each level~\cite[Sec.~4.6]{SPA3}.

The endpoint values of the fractional mean are
\begin{equation}
 \lim_{\lambda\to0}\bigl(\mathbb E \mathcal O(\tilde z)^\lambda\bigr)^{1/\lambda}
 =\exp\!\left[\mathbb E\log \mathcal O(\tilde z)\right],\qquad
 \lim_{\lambda\to1}\bigl(\mathbb E \mathcal O(\tilde z)^\lambda\bigr)^{1/\lambda}
 =\mathbb E \mathcal O(\tilde z).
 \label{eq:geometric-mean}
\end{equation}
The monotonicity of power means and continuity between these limits connect the geometric and arithmetic means~\cite{REMUniversality}. The exponent common to the observables is that of the masses;~\eqref{eq:meanidentity} allows it to be used in both cavity contributions.

\section{Constructive cavity and functional}
\label{sec:cavity}
The cavity formula compares two volumes using the same reservoir in the two contributions~\cite{AizenmanSimsStarrOriginal,AizenmanSimsStarr}. The decomposition of the fresh fields over the states of Section~\ref{sec:hierarchy} determines their hierarchical covariances. We then calculate the two averages using~\eqref{eq:meanidentity}, retaining the normalization and the network's bond weights.

\subsection{Cavity fields and identification with the ROSt}
\label{subsec:cavity-fields}
We start from the Aizenman--Sims--Starr formula~\cite{AizenmanSimsStarrOriginal,AizenmanSimsStarr}, in the convention where the temperature is already included in the fields:
\begin{equation}
 A(\mu)=\log2+\log\Bigl\langle\cosh\tilde x(\sigma_V)\Bigr\rangle_{\mu}
 -\log\Bigl\langle e^{\tilde y(\sigma_V)}\Bigr\rangle_{\mu}.
 \label{eq:cavity-functional}
\end{equation}

We consider a growing family of networks in which, before alignment, the existing unnormalized couplings retain the same law as the volume grows. The normalization is increasing. In the comparison between volumes, the common reservoir has Hamiltonian
\begin{equation}
 \frac{g(V^2)}{g((V\cup\{N+1\})^2)}H(\sigma_V).
 \label{eq:common-reservoir}
\end{equation}
Its Gibbs measure is that of~\eqref{eq:gibbs}, with the inverse temperature multiplied by the ratio in~\eqref{eq:common-reservoir}; this same measure is used in the two averages in~\eqref{eq:cavity-functional}.

In the aligned coordinates of Subsection~\ref{subsec:defects}, we define two families of fresh standard Gaussian variables, independent of each other and of the reservoir. Gaussian symmetry preserves this joint law even conditionally on the choice of ground state. The linear field is
\begin{equation}
 \tilde x(\sigma_V)=\frac{\beta}{g((V\cup\{N+1\})^2)}
 \sum_{i\in V}\sqrt{\operatorname{Var}(J_{N+1,i})+\operatorname{Var}(J_{i,N+1})}\,\tilde J_i\sigma_i.
 \label{eq:cavity-linear}
\end{equation}
For the couplings of the new site, the variance in~\eqref{eq:variance} is calculated with the normalization of the extended system. The variances of the old couplings still refer to the reservoir volume. The corrective contribution is
\begin{equation}
 \tilde y(\sigma_V)=\beta\sqrt{\frac1{g(V^2)^2}-\frac1{g((V\cup\{N+1\})^2)^2}}
 \sum_{i,j\in V}\sqrt{\operatorname{Var}(J_{ij})}\,\tilde J_{ij}\sigma_i\sigma_j.
 \label{eq:cavity-correction}
\end{equation}
We consider families for which the variances of the two fields remain bounded. The first collects the couplings incident on the new spin. The second compensates for the difference in the original coupling variances between the two normalizations. The correction concerns all pairs, including the cores.

Integrating out the new spin gives the average of the hyperbolic cosine in~\eqref{eq:cavity-functional}, together with the factor of two. Adding the corrective noise to the reservoir reconstructs the Hamiltonian of the old volume in distribution; by Gaussian symmetry, the corrective contribution is written with a positive sign in the exponent, as in the same formula. We divide the two partition functions by that of the reservoir: the disorder average of the difference of their logarithms therefore equals the mean cavity contribution. Any new diagonal coupling is centered and contributes zero to the average. Summing the increments along the growing family cancels the intermediate logarithms; the sum divided by the volume yields the thermodynamic pressure. The average of the increments retains the differences between the added sites, which may have different fields in a heterogeneous network. 

To pass to the IO description, keep each site's layer fixed along the growing family, independently of the disorder, with the limiting fractions of Eq.~\eqref{eq:fractions}. The same cavity identity then telescopes for the IO model when same-layer bonds are omitted from both the common reservoir and the fresh fields. At this fixed IO Gibbs measure, Gaussian integration by parts bounds the change in each expected logarithm upon restoring the omitted fresh terms by half their total variance. At fixed temperature, under the normalization condition following Eq.~\eqref{eq:core}, the limiting mean error per cavity contribution is bounded by
\begin{equation}
 C\beta^2\sum_{\ell\in\Lambda}p_\ell^2
 \leq C\beta^2\max_{\ell\in\Lambda}p_\ell
 \longrightarrow0.
 \label{eq:core-cavity-error}
\end{equation}
For the linear field, this follows by counting earlier sites in the added site's layer and averaging over additions. For the corrective field, summation by parts bounds the differences of inverse squared normalizations, weighted by squared volumes, by a constant times the final volume. Equation~\eqref{eq:core} then transfers the resulting averaged pressure from IO to the full model after refinement.

The partition of the sites into individual layers and of the pairs into blankets decomposes~\eqref{eq:cavity-linear}--\eqref{eq:cavity-correction} exactly into the constructive contributions of Ref.~\cite{SPA3}. They are evaluated on the same configurations that carry the masses~\eqref{eq:cascade}.

At fixed configurations, averaging over the fresh noise alone gives
\begin{equation}
\mathbb E\!\left[\tilde x(\sigma_V)\tilde x(\tau_V)\right]
 =\sum_{i\in V}(\Gamma_{N+1,i}+\Gamma_{i,N+1})\sigma_i\tau_i.
 \label{eq:linear-covariance}
\end{equation}
For the correction, we obtain
\begin{equation}
 \mathbb E\!\left[\tilde y(\sigma_V)\tilde y(\tau_V)\right]
 =\left[1-\frac{g(V^2)^2}{g((V\cup\{N+1\})^2)^2}\right]
 \sum_{i,j\in V}\Gamma_{ij}\sigma_i\tau_i\sigma_j\tau_j.
 \label{eq:quadratic-covariance}
\end{equation}
The linear and corrective fields are independent. These two formulas retain the entire variance matrix.

In the hierarchical representative, the product of the spins equals one on the shared layers and has mean equal to the square of the magnetization on the others. The weighted bounds~\eqref{eq:appendix-weighted-linear} and~\eqref{eq:appendix-weighted-quadratic} allow this substitution in~\eqref{eq:linear-covariance}--\eqref{eq:quadratic-covariance}. The difference between two consecutive levels of the linear covariance is
\begin{equation}
 (1-m_\ell^2)\sum_{i\in V_\ell}(\Gamma_{N+1,i}+\Gamma_{i,N+1}).
 \label{eq:linear-variance-increment}
\end{equation}
The common field is the average of the field over the fiber; its variance is
\begin{equation}
\mathbb E\!\left[\left\langle\tilde x(\sigma_V)\right\rangle_{m_\Lambda}^{\!2}\right]
 =\sum_{\ell\in\Lambda}m_\ell^2\sum_{i\in V_\ell}(\Gamma_{N+1,i}+\Gamma_{i,N+1}).
 \label{eq:common-linear-variance}
\end{equation}
The common linear field remains in the hyperbolic cosine and is integrated after the level averages. For the correction, the common field is the corresponding average over the fiber. Its diagonal part is common to all states; for the other terms, we also use the finite shell identity
\begin{equation}
 \left\langle\sigma_i\sigma_j\right\rangle_{m_\ell}
 =\frac{N_\ell m_\ell^2-1}{N_\ell-1},\qquad i\ne j,\qquad i,j\in V_\ell.
 \label{eq:shell-pair}
\end{equation}
For bounded total field variances, the deviation produced by these corrections is bounded above by a constant multiplied by the maximum of the inverses of the layer cardinalities. It therefore vanishes in the first limit, with macroscopic layers.

The limiting hierarchical covariances increase with the shared prefix. Conditional on the reservoir kernel and its complete masses, the two fresh families are jointly Gaussian and have zero mutual covariance. The weighted bounds identify their joint law on every finite collection of the same mass-carrying states. They also make the cross covariances between each shell mean and its residual vanish in the indicated limits: the corrective field uses the bounded-differences argument of Eq.~\eqref{eq:appendix-weighted-quadratic}, with the shell pair means as additional bounded weights. At fixed profile, Gaussianity then gives common components and node increments independent of the masses and of one another, with descendants reusing each ancestor's increment.

After the ordinary average over thermal continuations, finitely many complete condensed masses capture arbitrarily nearly all the measure. Bounded field variances bound the expected omitted contribution by a constant times its total mass, extending the joint identification to the complete sums. For the logarithms, the hyperbolic cosine is at least one; Jensen's inequality controls the negative logarithmic part of the exponential contribution through the weighted mean field, while uniform Gaussian moments control the positive parts. These bounds permit passage to the expected cavity functional, completing the weighted version of the constructive ROSt identification of Ref.~\cite[Sec.~4.5]{SPA3}.

\subsection{Evaluation of the functional}
\label{subsec:functional}
We evaluate Eq.~\eqref{eq:cavity-functional} on the masses in Eq.~\eqref{eq:cascade} of the common IO reservoir, obtained by applying the normalization in Eq.~\eqref{eq:common-reservoir} to the Hamiltonian in Eq.~\eqref{eq:IO}. The magnetizations and exponents are therefore those of the reservoir at the rescaled temperature; the coefficients of the fresh fields remain those in Eqs.~\eqref{eq:cavity-linear}--\eqref{eq:cavity-correction}. For the linear field, we represent the increments along a path by independent standard Gaussians. We collect the marks that remain to be integrated in the set
\begin{equation}
 \tilde z_{\Lambda_\ell}=\{\tilde z_t:t\in\Lambda_\ell\}.
 \label{eq:gaussian-profile}
\end{equation}
They are independent of the common field, represented by the fresh coefficients collected in the set
\begin{equation}
 \tilde J_V=\{\tilde J_i:i\in V\}.
 \label{eq:common-noise-profile}
\end{equation}
The variances~\eqref{eq:linear-variance-increment} and~\eqref{eq:common-linear-variance} yield the terminal function of the recursive variables in Ref.~\cite{SPA3}
\begin{equation}
\begin{aligned}
 &Y_L(\tilde J_V,\tilde z_\Lambda\mid m_\Lambda)\\
 &\quad=\cosh\!\Bigg[\left\langle\tilde x(\sigma_V)\right\rangle_{m_\Lambda}
 +\sum_{\ell\in\Lambda}\tilde z_\ell
 \sqrt{(1-m_\ell^2)\sum_{i\in V_\ell}(\Gamma_{N+1,i}+\Gamma_{i,N+1})}\,\Bigg].
\end{aligned}
 \label{eq:terminal}
\end{equation}
All network data and the partition remain fixed throughout the recursion. At each step, the exponent argument contains only the levels already integrated. Arguments indexed by an empty set are omitted. The average integrates only the new mark, with the previous data and the common field fixed:
\begin{equation}
 Y_{\ell-1}(\tilde J_V,\tilde z_{\Lambda_{\ell-1}}\mid m_\Lambda,\lambda_{\Lambda\setminus\Lambda_{\ell-1}})
 =\left[\mathbb E_\ell Y_\ell(\tilde J_V,\tilde z_{\Lambda_\ell}\mid m_\Lambda,\lambda_{\Lambda\setminus\Lambda_\ell})^{\lambda_\ell}\right]^{1/\lambda_\ell}.
 \label{eq:recursion}
\end{equation}
The index runs through the layers from last to first. At zero exponent, the recursion is understood through the geometric-mean limit~\eqref{eq:geometric-mean}.
The result obtained by iterating Eq.~\eqref{eq:recursion} from the terminal function~\eqref{eq:terminal} depends only on the common field. Equation~\eqref{eq:meanidentity} gives
\begin{equation}
 \mathbb E\log\left\langle\cosh\tilde x(\sigma_V)\right\rangle_{\bar\mu_{\mathrm{GREM}}}
 =\mathbb E\log Y_0(\tilde J_V\mid m_\Lambda,\lambda_\Lambda).
 \label{eq:linear-cavity-mean}
\end{equation}
The average on the right integrates the remaining common field.

For the correction, the Gaussian average of the exponential adds, at each level, half the variance of the increment multiplied by the exponent. The common field separates from the logarithm and has zero mean. The passage from the previous level to the current one replaces the square of the layer magnetization by one. In Eq.~\eqref{eq:quadratic-covariance}, the pairs within that layer and the pairs connecting it to the previous and subsequent layers therefore change. This yields the explicit functional
\begin{equation}
\begin{aligned}
 &\mathbb E A(\bar\mu_{\mathrm{GREM}})=\log2+
 \mathbb E\log Y_0(\tilde J_V\mid m_\Lambda,\lambda_\Lambda)\\
 &\quad-\frac12\left[1-\frac{g(V^2)^2}{g((V\cup\{N+1\})^2)^2}\right]
 \sum_{\ell\in\Lambda}\lambda_\ell(1-m_\ell^2)\Bigg[
 (1+m_\ell^2)\sum_{i,j\in V_\ell,\;i\ne j}\Gamma_{ij}\\
 &\qquad+\sum_{t<\ell}\;\sum_{i\in V_\ell,j\in V_t}(\Gamma_{ij}+\Gamma_{ji})
 +\sum_{t>\ell}m_t^2\sum_{i\in V_\ell,j\in V_t}(\Gamma_{ij}+\Gamma_{ji})\Bigg].
\end{aligned}
 \label{eq:functional}
\end{equation}
This is the formula for the hierarchical representative after the limit of macroscopic shells; the discrepancies are those controlled in the preceding sections. The diagonal produces no increments between states and remains in the common field. The temperature is already included in the variances and in the terminal function.

Equation~\eqref{eq:functional} is the rule for computing each contribution in the growing family of networks. The average along the growing IO family, followed by refinement and the comparison in Subsection~\ref{subsec:cavity-fields}, connects this evaluation to the pressure within the low-temperature description. A finer partition allows the magnetizations of the subsets to be distinguished while retaining the same sums over sites. The exponents and profiles representing the physical Gibbs measure are selected through the functional according to the model under consideration. The following examples make this rule and its fullRSB limit explicit.

\section{Examples}
\label{sec:examples}
The following examples show how the Blanket construction translates different interaction structures into explicit cavity calculations. Curie--Weiss provides the deterministic case, where the magnetization shells suffice to evaluate the energy and entropy. Introducing disorder leads to the hierarchical averages: for SK these recover the Parisi functional, while in restricted and deep Boltzmann machines they incorporate the fields generated by neighboring network layers. Under the conditions of Eq.~\eqref{eq:comparison}, the pressure comparison transfers the limiting mean pressures of the disordered models to their counterparts with non-Gaussian couplings.

\subsection{Curie--Weiss model}
\label{subsec:CW}
As in Ref.~\cite[Sec.~3.3]{BlanketTheory}, we consider the fully connected network with constant interaction. The coefficients and normalization are
\begin{equation}
 J_{ij}=J_0,\qquad H_{ij}=\frac{\beta J_0}{N},\qquad g(V^2)=N.
 \label{eq:CW-coefficients}
\end{equation}
This deterministic case of the model~\eqref{eq:model} explicitly retains the mean value of the couplings. The total magnetization is expressed in terms of the layer magnetizations:
\begin{equation}
 m(\sigma_V)=\frac1N\sum_{i\in V}\sigma_i
 =\sum_{\ell\in\Lambda}p_\ell m_\ell.
 \label{eq:CW-magnetization}
\end{equation}
The Hamiltonian depends only on this quantity:
\begin{equation}
 H(\sigma_V)=\beta NJ_0\,m(\sigma_V)^2.
 \label{eq:CW-Hamiltonian}
\end{equation}
We consider the ferromagnetic case, which, with the sign adopted in Eq.~\eqref{eq:gibbs}, corresponds to $J_0<0$. The reference configuration with all spins positive is already a ground state.

Equation~\eqref{eq:core-local} gives, for the core,
\begin{equation}
 \widehat H_\ell(\sigma_{V_\ell})=\beta NJ_0p_\ell^2m_\ell^2,
 \label{eq:CW-core}
\end{equation}
and Eq.~\eqref{eq:interface-local} gives, for the interface, counting both orientations,
\begin{equation}
 \overline H_\ell(\sigma_{V_\ell}\mid\sigma_{Q_{\ell-1}})
 =2\beta NJ_0p_\ell m_\ell\sum_{t<\ell}p_t m_t.
 \label{eq:CW-interface}
\end{equation}
The deterministic bound on the core is therefore
\begin{equation}
 \frac1N\sup_{\sigma_V}|H(\sigma_V)-\bar H(\sigma_V)|
 \leq\beta|J_0|\sum_{\ell\in\Lambda}p_\ell^2\longrightarrow0.
 \label{eq:CW-core-bound}
\end{equation}
The limit follows from refinement of the partition. On every shell with fixed total magnetization, the energy is constant and the Gibbs measure is uniform. The binomial multiplicity of the shell determines the pressure through
\begin{equation}
 -\lim_{N\to\infty}f=\max_{-1\leq m\leq1}\left\{-\frac{1+m}{2}\log\frac{1+m}{2}
 -\frac{1-m}{2}\log\frac{1-m}{2}-\beta J_0m^2\right\}.
 \label{eq:CW-entropy-functional}
\end{equation}
Here $m$ is the value of the total magnetization on the shell; the entropic terms at the endpoints are understood by continuity. The formula follows from the disintegration of the partition function by magnetization and from Stirling's formula: the number of shells grows only linearly with the volume.

The same pressure is obtained from the cavity. The common reservoir in Eq.~\eqref{eq:common-reservoir} has the normalization of the extended system. The deterministic contributions, denoted by $x$ and $y$, are respectively
\begin{equation}
 x(\sigma_V)=\frac{2\beta J_0}{N+1}\sum_{i\in V}\sigma_i,\qquad
 y(\sigma_V)=-\frac{\beta J_0}{N(N+1)}
 \left(\sum_{i\in V}\sigma_i\right)^2.
 \label{eq:CW-cavity-fields}
\end{equation}
The first collects the couplings of the new spin. The second is the negative of the difference between the energy of the old system and that of the common reservoir; it therefore enters with a positive sign in the exponent in the second term of Eq.~\eqref{eq:cavity-functional}. In that formula, these two contributions replace the Gaussian fields, with the same reservoir Gibbs measure in both averages. Dividing both partition functions by that of the reservoir gives the exact identity
\begin{equation}
 \log\frac{\sum_{\sigma_{V\cup\{N+1\}}}e^{-H(\sigma_{V\cup\{N+1\}})}}
 {\sum_{\sigma_V}e^{-H(\sigma_V)}}
 =A(\mu)-\frac{\beta J_0}{N+1}.
 \label{eq:CW-cavity-increment}
\end{equation}
The last term comes from the new diagonal coupling and is negligible in the limit. The sum of the increments, divided by the volume, yields the pressure.

For a thermodynamic magnetization selected by Eq.~\eqref{eq:CW-entropy-functional}, the limit of the contribution is
\begin{equation}
 \lim_{N\to\infty}A(\mu)
 =\log\!\left[2\cosh(2\beta J_0m)\right]+\beta J_0m^2.
 \label{eq:CW-cavity-functional}
\end{equation}
The stationarity condition for Eq.~\eqref{eq:CW-entropy-functional} gives
\begin{equation}
 m=-\tanh(2\beta J_0m).
 \label{eq:CW-self-consistency}
\end{equation}
Substituting this relation shows that, at the thermodynamic maxima, the entropic value coincides with that of Eq.~\eqref{eq:CW-cavity-functional}. In the ordered ferromagnetic regime, the two opposite magnetizations give the same contribution, since both cavity observables are even. The constant case thus shows the reduction of CCM to the computation of the magnetization and the deterministic correction, whereas in the following examples the dispersion of energies requires hierarchical averages.

\subsection{Sherrington--Kirkpatrick model}
For SK~\cite{SherringtonKirkpatrick,SPA3}, in the Gaussian representative the ordered couplings are independent standard Gaussians before alignment. The choices of the general parameters are
\begin{equation}
 \operatorname{Var}(J_{ij})=1,\qquad H_{ij}=\frac{\beta J_{ij}}{\sqrt N},\qquad g(V^2)=\sqrt N.
 \label{eq:SK-normalization}
\end{equation}
Thus
\begin{equation}
 H(\sigma_V)=\frac{\beta}{\sqrt N}\sum_{i,j\in V}J_{ij}\sigma_i\sigma_j.
 \label{eq:SK-model}
\end{equation}
Equations~\eqref{eq:cavity-linear}--\eqref{eq:cavity-correction} give, at finite volume,
\begin{equation}
 \tilde x(\sigma_V)=\beta\sqrt{\frac2{N+1}}\sum_{i\in V}\tilde J_i\sigma_i,\qquad
 \tilde y(\sigma_V)=\frac{\beta}{\sqrt{N(N+1)}}\sum_{i,j\in V}\tilde J_{ij}\sigma_i\sigma_j.
 \label{eq:SK-cavity-fields}
\end{equation}
The noises in the two families are fresh standard Gaussians, independent of one another and of the reservoir. For two states sharing the prefix up to layer $\ell$, the finite-volume covariances and the values of the hierarchical representative are
\begin{equation}
 \mathbb E\!\left[\tilde x(\sigma_V)\tilde x(\tau_V)\right]
 =\frac{2\beta^2}{N+1}\sum_{i\in V}\sigma_i\tau_i
 \longrightarrow2\beta^2\rho_\ell,
 \label{eq:SK-linear-covariance}
\end{equation}
\begin{equation}
 \mathbb E\!\left[\tilde y(\sigma_V)\tilde y(\tau_V)\right]
 =\frac{\beta^2}{N(N+1)}\left(\sum_{i\in V}\sigma_i\tau_i\right)^2
 \longrightarrow\beta^2\rho_\ell^2.
 \label{eq:SK-quadratic-covariance}
\end{equation}
Replacing the overlaps by the profile~\eqref{eq:rho} uses the estimates and the order of limits in subsection~\ref{subsec:hierarchy}. The initial value describes the common component; consecutive differences give the variances of the increments. We collect the profile as
\begin{equation}
 \rho_\Lambda=\{\rho_\ell:\ell\in\Lambda\},\qquad
 0\leq\rho_0\leq\rho_1\leq\cdots\leq\rho_L=1.
 \label{eq:SK-profile}
\end{equation}
We represent the common field by a fresh standard Gaussian $\tilde z$, independent of the marks $\tilde z_\ell$. The terminal function contains both components, with amplitudes determined by Eq.~\eqref{eq:SK-linear-covariance}:
\begin{equation}
 Y_L(\tilde z,\tilde z_\Lambda\mid\rho_0,\rho_\Lambda)
 =\cosh\!\left[\beta\sqrt{2\rho_0}\,\tilde z+
 \beta\sum_{\ell\in\Lambda}\sqrt{2(\rho_\ell-\rho_{\ell-1})}\,\tilde z_\ell\right].
 \label{eq:SK-terminal}
\end{equation}
The recursion integrates the marks from the last level to the first:
\begin{equation}
 Y_{\ell-1}(\tilde z,\tilde z_{\Lambda_{\ell-1}}\mid\rho_0,\rho_\Lambda,\lambda_{\Lambda\setminus\Lambda_{\ell-1}})
 =\left[\mathbb E_\ell
 Y_\ell(\tilde z,\tilde z_{\Lambda_\ell}\mid\rho_0,\rho_\Lambda,\lambda_{\Lambda\setminus\Lambda_\ell})^{\lambda_\ell}\right]^{1/\lambda_\ell}.
 \label{eq:SK-recursion}
\end{equation}
The step with zero exponent uses the geometric-mean limit~\eqref{eq:geometric-mean}. The final average integrates only $\tilde z$. For the correction, the identity~\eqref{eq:meanidentity} assigns to each level half the increment of the covariance~\eqref{eq:SK-quadratic-covariance}, multiplied by the exponent of the level. The common component of the correction has zero mean. This gives
\begin{equation}
 \mathbb E A(\bar\mu_{\mathrm{GREM}})=\log2+ \mathbb E\log Y_0(\tilde z\mid \rho_0,\rho_\Lambda,\lambda_\Lambda) -\frac{\beta^2}{2}\sum_{\ell\in\Lambda}\lambda_\ell (\rho_\ell^2-\rho_{\ell-1}^2).
 \label{eq:SK-functional}
\end{equation}
For SK, the thermodynamic pressure is the infimum of Eq.~\eqref{eq:SK-functional} over the profiles~\eqref{eq:SK-profile}, nondecreasing exponents in $[0,1]$, and finite depths~\cite{ParisiOrderParameter,GuerraRSBBound,TalagrandParisiFormula}.

\subsection{Restricted and deep Boltzmann machines}
We now introduce the network architecture~\cite{CCMBoltzmannMachines,ChenIssaMourrat2026,BoursierMourrat2026}. The sites are divided into $K$ disjoint layers $V^{(a)}$, indexed by
\begin{equation}
 a\in A:=\{1,\ldots,K\}.
 \label{eq:DBM-layer-indices}
\end{equation}
Their proportions satisfy
\begin{equation}
 |V^{(a)}|=\lambda_a N+o(N),\qquad
 \lambda_a>0,\qquad \sum_{a\in A}\lambda_a=1.
 \label{eq:DBM-proportions}
\end{equation}
Each site interacts only with the adjacent layers. Two layers give the bipartite RBM; three or more layers give the DBM. We denote by $\beta_a$ the strength, including the temperature, of the couplings between $a$ and $a+1$. For each orientation of a pair between these layers, the choices of the general parameters are
\begin{equation}
 \operatorname{Var}(J_{ij})=\left(\frac{\beta_a}{\beta}\right)^2,\qquad
 H_{ij}=\frac{\beta J_{ij}}{\sqrt N},\qquad g(V^2)=\sqrt N.
 \label{eq:DBM-coefficients}
\end{equation}
The ratio between the strengths and the inverse temperature determines the variance of the unnormalized couplings. Before alignment, the ordered couplings in the Gaussian representative are independent and centered; the two orientations combine into a single standard noise per pair:
\begin{equation}
 H(\sigma_V)\stackrel d=\sqrt{\frac2N}\sum_{a<K}\beta_a
 \sum_{i\in V^{(a)}}\sum_{j\in V^{(a+1)}}J_{ij}^{(a)}\sigma_i\sigma_j.
 \label{eq:DBM-model}
\end{equation}
\Needspace{3\baselineskip}
The Blanket partition cuts across the architectural layers. Figure~\ref{fig:DBM} shows the subdivision on the graph and in the matrix.\par
\afterpage{\clearpage
\begin{figure}[p]
\centering
\includegraphics[width=\textwidth,height=.62\textheight,keepaspectratio]{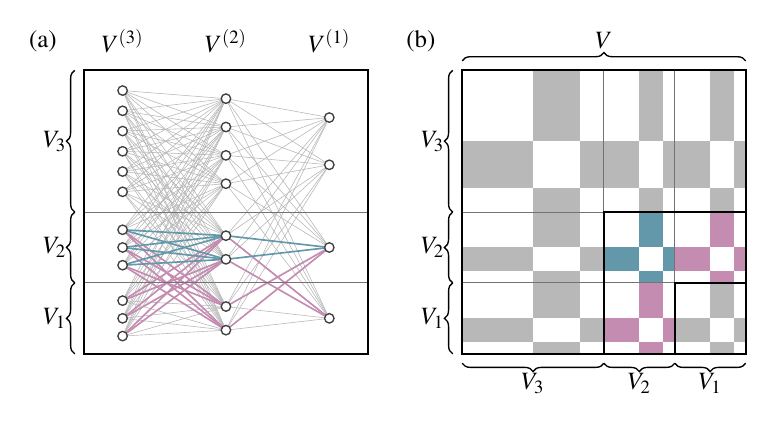}
\begingroup\small\setstretch{1}
\caption{Blanket partition of a DBM, on the graph and in the interaction matrix~\cite{CCMBoltzmannMachines}. The architectural layers $V^{(a)}$ are coupled only to their neighbors; sectors without interactions remain white. Each layer $V_\ell$ of the partition brings together the cells cut out within the different architectural layers. Blue identifies the cores and magenta the interfaces, with the same colors in the matrix and in the graph. The sizes of the cells and pair sectors are those in Eqs.~\eqref{eq:DBM-cell-count} and~\eqref{eq:DBM-edge-count}; the cell magnetizations then determine the profiles~\eqref{eq:DBM-physical-profile} used in the fields.}
\label{fig:DBM}
\endgroup
\end{figure}
\clearpage}
In the ordered representation, the variances of the adjacent couplings are
\begin{equation}
 \Gamma_{ij}=\Gamma_{ji}=\frac{\beta_a^2}{N},\qquad
 i\in V^{(a)},\qquad j\in V^{(a+1)}.
 \label{eq:DBM-bond-variance}
\end{equation}
The other coefficients vanish. We study the pressure per spin at fixed $K$, proportions, and strengths, computing its CCM functional at prescribed profiles.

In each architectural layer, we choose the nested sets given by its intersections with the cumulative sets of the partition and define
\begin{equation}
 q_\ell^{(a)}=\frac{|Q_\ell\cap V^{(a)}|}{|V^{(a)}|},
 \qquad q_0^{(a)}=0,\qquad q_L^{(a)}=1.
 \label{eq:DBM-geometric-profile}
\end{equation}
Each layer of the partition brings together the cells obtained by intersecting it with the layers of the network; the size of each cell is
\begin{equation}
 |V_\ell\cap V^{(a)}|=|V^{(a)}|
 (q_\ell^{(a)}-q_{\ell-1}^{(a)}).
 \label{eq:DBM-cell-count}
\end{equation}
The superscript $a$ denotes an architectural layer of the network; the subscript $\ell$ denotes a layer of the Blanket partition.

 In the rectangle between two adjacent layers, we have
\begin{equation}
\begin{aligned}
 &W_\ell\cap(V^{(a)}\times V^{(a+1)})
 =[(V_\ell\cap V^{(a)})\times(Q_{\ell-1}\cap V^{(a+1)})]\\
 &\quad\cup[(Q_{\ell-1}\cap V^{(a)})\times(V_\ell\cap V^{(a+1)})]
 \cup[(V_\ell\cap V^{(a)})\times(V_\ell\cap V^{(a+1)})].
\end{aligned}
 \label{eq:DBM-edge-partition}
\end{equation}
The first two rectangles are the interfaces; the third is the core. The second orientation occupies the transposed rectangle. The number of pairs in the blanket is
\begin{equation}
 |W_\ell\cap(V^{(a)}\times V^{(a+1)})|
 =|V^{(a)}|\,|V^{(a+1)}|
 (q_\ell^{(a)}q_\ell^{(a+1)}-q_{\ell-1}^{(a)}q_{\ell-1}^{(a+1)}).
 \label{eq:DBM-edge-count}
\end{equation}
Refinement controls the removal of the core from the energy; the correction field retains the full count of pairs.

To apply the general construction, the partition is first refined into cells, keeping the nonempty ones macroscopic in the first thermodynamic limit; the steps with the same exponent are then combined. Disintegration into shells fixes the rectified magnetization of each nonempty cell:
\begin{equation}
 m_\ell^{(a)}=\frac1{|V_\ell\cap V^{(a)}|}
 \sum_{i\in V_\ell\cap V^{(a)}}\sigma_i.
 \label{eq:DBM-cell-magnetization}
\end{equation}
Sums over cells include only the nonempty ones. The profile of the physical correlations in layer $a$ is
\begin{equation}
 \rho_\ell^{(a)}=q_\ell^{(a)}+
 \sum_{t>\ell}(q_t^{(a)}-q_{t-1}^{(a)})(m_t^{(a)})^2,
 \qquad 0\leq\ell\leq L.
 \label{eq:DBM-physical-profile}
\end{equation}
The initial value is the common datum and the terminal value is one. As in Eq.~\eqref{eq:rho-increments}, each step adds the relative volume of the cell multiplied by the variance of its spin. The complete profiles are
\begin{equation}
 \rho_{\Lambda\cup\{0\}}^{(A)}=\{\rho_\ell^{(a)}:a\in A,\ \ell\in\Lambda\cup\{0\}\}.
 \label{eq:DBM-profiles}
\end{equation}
To distinguish the hierarchical exponents from the architectural proportions~\eqref{eq:DBM-proportions}, we denote the former by $\zeta_\ell$ and their collection by $\zeta_\Lambda$.

The fresh field of a spin inserted into layer $a$ is obtained from Eq.~\eqref{eq:cavity-linear}. At finite volume, it is
\begin{equation}
 \tilde x_a(\sigma_V)=\sqrt{\frac2{N+1}}\left[
 \beta_{a-1}\sum_{i\in V^{(a-1)}}\tilde J_i^{(a)}\sigma_i+
 \beta_a\sum_{i\in V^{(a+1)}}\tilde J_i^{(a)}\sigma_i\right].
 \label{eq:DBM-field}
\end{equation}
The noises are standard Gaussians, independent and fresh; at the boundaries, the missing neighbor is omitted. The correction field in Eq.~\eqref{eq:cavity-correction}, on the other hand, includes all existing adjacent couplings:
\begin{equation}
 \tilde y(\sigma_V)=\sqrt{\frac2{N(N+1)}}\sum_{a<K}\beta_a
 \sum_{i\in V^{(a)}}\sum_{j\in V^{(a+1)}}\tilde J_{ij}^{(a)}\sigma_i\sigma_j.
 \label{eq:DBM-cavity-correction}
\end{equation}
These noises are fresh standard Gaussians, independent of those of the linear field and of the reservoir. The correction also includes the pairs within the cores and is common to all choices of the layer to which the site is added. Both cavity averages use the same reservoir Gibbs measure.

We now apply the hierarchical representative of subsection~\ref{subsec:hierarchy}, with the same estimates and the same order of limits. For states sharing the first $\ell$ layers of the partition, the mean product of spins in each architectural layer is given by the profile~\eqref{eq:DBM-physical-profile}. The finite-volume covariance of the linear field and its limit are
\begin{equation}
 \mathbb E\!\left[\tilde x_a(\sigma_V)\tilde x_a(\tau_V)\right]
 =\frac2{N+1}\left[\beta_{a-1}^2\sum_{i\in V^{(a-1)}}\sigma_i\tau_i
 +\beta_a^2\sum_{i\in V^{(a+1)}}\sigma_i\tau_i\right]\longrightarrow Q_\ell^{(a)},
 \label{eq:DBM-linear-covariance}
\end{equation}
where
\begin{equation}
 Q_\ell^{(a)}=2\beta_{a-1}^2\lambda_{a-1}\rho_\ell^{(a-1)}
 +2\beta_a^2\lambda_{a+1}\rho_\ell^{(a+1)}.
 \label{eq:DBM-field-variance}
\end{equation}
The superscript distinguishes this variance from the set of sites $Q_\ell$. For the same states, the covariance of the correction field is
\begin{equation}
\begin{aligned}
 &\mathbb E\!\left[\tilde y(\sigma_V)\tilde y(\tau_V)\right]
 =\frac2{N(N+1)}\sum_{a<K}\beta_a^2
 \left(\sum_{i\in V^{(a)}}\sigma_i\tau_i\right)
 \left(\sum_{j\in V^{(a+1)}}\sigma_j\tau_j\right)\\
 &\qquad\longrightarrow2\sum_{a<K}\beta_a^2\lambda_a\lambda_{a+1}\rho_\ell^{(a)}\rho_\ell^{(a+1)}.
\end{aligned}
 \label{eq:DBM-correction-covariance}
\end{equation}
The initial value of each covariance is the variance of the common datum; successive differences are the variances of the increments. For each layer, we represent the common linear field by a fresh standard Gaussian $\tilde z$, independent of the level marks, collected as in Eq.~\eqref{eq:gaussian-profile}. The profile argument of each recursive variable is restricted to the adjacent architectural layers. The terminal function is
\begin{equation}
 Y_L^{(a)}(\tilde z,\tilde z_\Lambda^{(a)}\mid\rho_{\Lambda\cup\{0\}}^{(A\cap\{a-1,a+1\})})
 =\cosh\!\left[\sqrt{Q_0^{(a)}}\,\tilde z+
 \sum_{\ell\in\Lambda}\sqrt{Q_\ell^{(a)}-Q_{\ell-1}^{(a)}}\,\tilde z_\ell^{(a)}\right].
 \label{eq:DBM-terminal}
\end{equation}
Each new mark is then integrated while keeping the prefix and the common field fixed:
\begin{equation}
\begin{aligned}
 &Y_{\ell-1}^{(a)}(\tilde z,\tilde z_{\Lambda_{\ell-1}}^{(a)}\mid\rho_{\Lambda\cup\{0\}}^{(A\cap\{a-1,a+1\})},\zeta_{\Lambda\setminus\Lambda_{\ell-1}})\\
 &\quad=\left[\mathbb E_\ell
 Y_\ell^{(a)}(\tilde z,\tilde z_{\Lambda_\ell}^{(a)}\mid\rho_{\Lambda\cup\{0\}}^{(A\cap\{a-1,a+1\})},\zeta_{\Lambda\setminus\Lambda_\ell})^{\zeta_\ell}\right]^{1/\zeta_\ell}.
\end{aligned}
 \label{eq:DBM-recursion}
\end{equation}
Exponents zero and one give the geometric and arithmetic means in Eq.~\eqref{eq:geometric-mean}. The same exponents govern all layers because the masses and ancestors belong to a single hierarchy.

The identity~\eqref{eq:meanidentity} applied to the correction assigns to each level half the increment in Eq.~\eqref{eq:DBM-correction-covariance}, multiplied by the exponent of the level. The factor of two in the covariance thus cancels the factor of one half in the Gaussian average. We grow the network by adding sites to the different layers with limiting frequencies equal to their proportions. The average of the cavity increments then weights the linear contribution of each layer by the same proportion, while the correction appears once. We denote this average by $A(\mu)$. We obtain
\begin{equation}
\begin{aligned}
 \mathbb E A(\bar\mu_{\mathrm{GREM}})
 ={}&\log2+\sum_{a\in A}\lambda_a\mathbb E\log
 Y_0^{(a)}(\tilde z\mid
 \rho_{\Lambda\cup\{0\}}^{(A\cap\{a-1,a+1\})},\zeta_\Lambda)\\
 &-\sum_{a<K}\beta_a^2\lambda_a\lambda_{a+1}
 \sum_{\ell\in\Lambda}\zeta_\ell
 (\rho_\ell^{(a)}\rho_\ell^{(a+1)}-\rho_{\ell-1}^{(a)}\rho_{\ell-1}^{(a+1)}).
\end{aligned}
 \label{eq:DBM-functional}
\end{equation}
The common field is integrated last; the common correction has zero mean. The formula in Ref.~\cite{CCMBoltzmannMachines} is thus expressed in terms of the physical profiles.

For the fullRSB form, we fix continuous nondecreasing profiles satisfying
\begin{equation}
 0\leq\rho^{(a)}(t)\leq1,\qquad \rho^{(a)}(1)=1,\qquad 0\leq t\leq1.
 \label{eq:DBM-continuous-profiles}
\end{equation}
The initial values describe the common field. The exponent is nondecreasing and right-continuous, with
\begin{equation}
 0\leq\zeta(t)\leq1,\qquad 0\leq t\leq1.
 \label{eq:DBM-continuous-exponent}
\end{equation}
The finite recursion samples the path using
\begin{equation}
 \rho_\ell^{(a)}=\rho^{(a)}(\ell/L),\qquad 0\leq\ell\leq L;
 \qquad \zeta_\ell=\zeta((\ell-1)/L),\qquad \ell\in\Lambda.
 \label{eq:DBM-path-sampling}
\end{equation}
Each profile increment therefore uses the exponent at the left endpoint. The limit of Eq.~\eqref{eq:DBM-functional}, with the same initial data and the same discretizations in all terms, is
\begin{equation}
\begin{aligned}
 \lim_{L\to\infty}\mathbb E A(\bar\mu_{\mathrm{GREM}})
 ={}&\log2+\sum_{a\in A}\lambda_a
 \lim_{L\to\infty}\mathbb E\log
 Y_0^{(a)}(\tilde z\mid
 \rho_{\Lambda\cup\{0\}}^{(A\cap\{a-1,a+1\})},\zeta_\Lambda)\\
 &-\sum_{a<K}\beta_a^2\lambda_a\lambda_{a+1}
 \int_0^1\zeta(t)\,\mathrm d[\rho^{(a)}(t)\rho^{(a+1)}(t)].
\end{aligned}
 \label{eq:DBM-fullRSB}
\end{equation}
The limit of the linear contribution is defined by the terminal function~\eqref{eq:DBM-terminal} and the recursion~\eqref{eq:DBM-recursion}; the cost is the Stieltjes integral. Continuity of the profiles makes the value of the exponent at its jump points irrelevant.

\section{Conclusions}
We have shown how Blanket theory~\cite{SimplifiedParisi,RSBWithoutReplicas,REMUniversality,BlanketTheory,Multispin,SPA3} can be used to construct and calculate cavity functionals for dense spin networks. Combining its decomposition of the interactions with the low-temperature mechanism developed in Ref.~\cite{SPA3} leads to the explicit functional~\eqref{eq:functional}. The layer magnetizations and hierarchical exponents describe the states and their weights, while the interaction variances and network normalization determine the coefficients entering the calculation. Averaging the resulting cavity contributions along a growing family of networks gives the pressure of the corresponding hierarchical representative.

Control of the core and of the term bilinear in the defects connects the original Hamiltonian to this representation. Refining the partition suppresses the interactions internal to the layers, while dilution makes the bilinear remainder negligible under the bounds~\eqref{eq:core} and~\eqref{eq:bilinear}. The remaining linear energies describe fluctuations around the chosen ground state. Disintegration of the Gibbs measure into magnetization shells retains both their multiplicities and their energy centers, so the comparison between fibers includes the full energetic and entropic contributions. Sharing defects along an ancestral path and renewing their supports on different branches then provides the physical mechanism for the hierarchy. The population bounds allow this organization to persist for exponentially many states in the dilute regime.

Within this hierarchy, the regular energy edge determines the exponents of the masses through the slopes of the costs in~\eqref{eq:slope}. These costs use the effective law induced by ground-state rectification. The resulting Poisson representation yields the fractional-mean identity~\eqref{eq:meanidentity}; applying it successively from the last level to the first produces the recursive averages. Its geometric and arithmetic limits describe the endpoints of the same averaging rule, with exponents fixed by the masses and shared by both cavity observables. Thus the energy statistics supply the parameters needed to calculate the averages over the hierarchical Gibbs weights.

For heterogeneous networks, the essential extension is to control the geometry of the states with the coupling variances retained as weights. The uniform estimates in Section~\ref{sec:hierarchy} permit the same shared configurations to determine the covariances of the fresh fields, completing the constructive identification with the Aizenman--Sims--Starr ROSt. The linear contribution depends on the couplings incident on the added spin; the corrective contribution accounts for the change in normalization of all existing couplings. Their averages use the same reservoir Gibbs measure. The common linear field remains inside the terminal hyperbolic cosine and is integrated after the hierarchical increments, whereas the common corrective field contributes zero after the disorder average. These features determine both the recursion and the explicit cost in~\eqref{eq:functional}.

The examples show how this rule adapts to different interaction structures. In Curie--Weiss, the energy is constant on each magnetization shell, and the shell entropy and deterministic cavity calculation give the same thermodynamic pressure. For SK, the variances reduce to functions of the physical overlaps, recovering the canonical Parisi functional~\eqref{eq:SK-functional}. In RBM and DBM, intersecting the Blanket partition with the architectural layers gives the cell magnetizations and the distinct fields generated by neighboring layers. Their contributions combine into~\eqref{eq:DBM-functional}, whose fullRSB limit~\eqref{eq:DBM-fullRSB} retains a common hierarchy of exponents and a cost determined by adjacent pairs of layers.

The pressure comparison~\eqref{eq:comparison} extends the result to non-Gaussian couplings and is performed at fixed temperature before rectification. Under its moment and normalization conditions, the limiting pressure agrees with that of the Gaussian representative. The physical identification of the hierarchical states relies on the dilute regime and the energy-edge regularity described in Ref.~\cite{SPA3}, with the thermodynamic limit taken first on macroscopic layers. The functional thus encodes the thermodynamics of the network. Selecting the physical value becomes a mathematical problem: determining the stationary profiles under admissible variations, analysing their stability, and comparing the corresponding functional values. For SK, this selection is expressed by the established minimization principle.

\section*{Acknowledgments}\label{sec:acknowledgments}
We thank Francesco Concetti (Unidistance), Giampiero Bardella (Sapienza Universita di Roma), Nicola Kistler (Goethe University Frankfurt) and ChatGPT (OpenAI) for important discussions and suggestions.

\Needspace{8\baselineskip}
\section*{Declaration of Generative AI}
Research, writing, and proofreading of the present manuscript were performed with AI assistance (GPT-6 Astra) under strict human guidance and scrutiny. We independently checked all mathematical arguments and take full responsibility for the content of the publication.

\begingroup
\small
\setstretch{1.05}
\raggedright

\endgroup
\end{document}